\documentstyle[prl,aps,preprint]{revtex}

\begin{document}
\preprint{OUTP-96-11 S,~~
cond-mat/9602163 }
\title
{\bf The Liouville Theory as a Model for Prelocalized States in
Disordered Conductors}
\author{Ian. I. Kogan, C. Mudry$^*$  and A. M. Tsvelik}
\address{Department of  Physics, University of Oxford, 1 Keble Road, Oxford OX1 3NP,  UK\\
$^*$Department of  Physics, Massachusetts Institute of Technology, 77
Massachusetts Avenue, Cambridge, MA 02139, USA}
\maketitle
\begin{abstract}

It is established that the distribution of the zero energy
eigenfunctions of (2 + 1)-dimensional Dirac electrons in a random
gauge potential is described by the Liouville model. This model has a
line of critical points parameterized by the strength of
disorder and  the
scaling dimensions of the inverse participation ratios 
coincide with the  dimensions obtained  in  the conventional
localization theory. From this fact  we conclude that the renormalization group
trajectory of the latter theory lies in the vicinity of the line of
critical points of the Liouville model.

\end{abstract}

Among the problems of localization theory there is one which, 
until recently, had attracted less attention than it deserves. 
This is the problem
of spatial correlations of wave functions at distances much smaller
than the localization length $L_c$ (see Refs.\cite{wegner,alt}). 
This  problem is well understood only
for extended states, {\it i.e.}, in the limit 
of small wave function amplitudes 
$t = |\psi({\bf x})|^2$. Since extended states explore the entire sample  
one can neglect their spatial variations and treat  
the Hamiltonian as a random matrix. The  distribution function
$P(t)$ derived by the methods of random matrix theory depends only on
the global symmetry of the random ensemble and has an
approximately Gaussian form (the Porter-Thomas distribution; see for
example \cite{porter}). This 
approach fails for larger $t$'s 
since the tails of the distribution function are
determined by rare spatially inhomogeneous 
configurations with high local amplitudes. The first calculation 
for the asymptotic form of $P(t)$ in small two-dimensional samples ($L << L_c$)
was performed by Altshuler, Kravtsov and Lerner\cite{alt}  
using the renormalization
group and replicas. The  nonperturbative approach based on the
supersymmetric $\sigma$-model had not been used in this context until
Muzykantskii and
Khmelnitskii\cite{dima} pointed out that in order to describe  
exceptional events most affected by the disorder one should look for a
saddle point of the supersymmetric $\sigma$-model.  This idea 
has then been exploited by Fal'ko and Efetov \cite{falko} 
who have derived a reduced $\sigma$-model adapted to the studies of
properties of a single quantum state in the  discrete spectrum of a
confined system and found that the reduced $\sigma$-model has a
non-trivial vacuum. They have 
calculated $P(t)$ in two loops (the saddle point 
approximation with  Gaussian
fluctuations around it). 
 
 Below  we discuss only two-dimensional systems. In two-dimensions 
the conventional localization theory of unitary ensemble represents a 
special case since the first loop localization correction to the 
conductivity vanishes and the
 localization length is therefore very 
large: $L_c \sim \exp\left(G_0^2\right)$, 
where $G_0$ is the bare conductance (we assume that $G_0 >> 1$). Meanwhile 
the behaviour of wave functions 
becomes nontrivial at 
much smaller  length scales $L > \exp\left(G_0\right)$.  In this case the 
results of  Altshuler {\it et al.} and Fal'ko and Efetov suggest that 
the renormalization flow  for $P(t)$, 
yet eventually turning to the strong coupling
(localization), spends a lot of ``time'' in the vicinity of some 
critical line where the asymptotics  of $P(t)$  is given by the
log-normal distribution. It is difficult,
however, to identify this critical line within the replica
approach. The supersymmetric saddle point calculations provide a
better 
insight since they give  the Liouville equation as the
$\sigma$-model's saddle point condition. 

 In this Letter we describe a  
model of disorder where   the line of critical points discovered in
Refs. \cite{alt,falko} is stable. This is the theory of 
(2 + 1)-dimensional Dirac fermions in a random gauge potential (FRGP
model). We show that the distribution function of 
the prelocalized states in this
model is described by the Liouville field theory
(LFT). This theory  was introduced by Polyakov \cite{polyakov} 
in the context of string theory and has been extensively studied.  We
are going to use this  accumulated knowledge for the theory of localization. 
One important property of LFT is that it does have a stable line of critical
points which seems to describe the prelocalized states in the
conventional theory of localization. As we shall show later, this 
line is parameterized by the strength of disorder.

Until  present time only two  critical disordered systems have been
studied: the model of a half-filled Landau level 
(see, for example, \cite{jans} and references therein) 
and  the model of (2 + 1)-dimensional Dirac
fermions in a random gauge potential (FRGP model)
\cite{ludwig,ners,mudry,caux}. The latter  model has  an unbounded
energy 
spectrum, but the spectrum of $\hat H^2$ is bounded from below ($E_n^2
\geq 0$). Since the conductivity of the FRGP model at zero frequency is
finite 
and the wave functions with $E = 0$ have multifractal properties 
(see Ludwig {\it et al.} \cite{ludwig}), we suggest  that 
the mobility edge of this model coincides with  the boundary of 
the spectrum generating operator $\hat H^2$. This conjecture  does  not
 generally hold in more complex critical disordered systems (e.g. $2D$
 symplectic and $3D$).

In this letter we study the statistics of $E = 0$ wave functions of 
the Abelian FRGP model in two dimensions \cite{alter}. 
In the gauge where $A_{\mu} = \epsilon_{\mu\nu}\partial_{\nu}\Phi$ the 
Dirac equation has one obvious solution with $E = 0$, which we write down 
in the normalized  form:
\begin{equation}
\psi_{\sigma}({\bf x}) 
= (1 \pm \sigma^3)_{\sigma,\sigma'}
\frac{e^{- \sigma'\Phi({\bf x})}}{[\int \mbox{d}^2x' e^{-2\Phi({\bf x}')}]^{1/2}}.
\label{wave}
\end{equation}
This solution is unique on a closed sphere if the total flux is less than 2. 
For greater fluxes the index theorem predicts the existence of other 
solutions with $E = 0$ \cite{aharonov}. 

 We shall consider  $\nabla\Phi$ as a random variable with 
the Gaussian distribution:
\begin{equation}
P[\Phi] = Z_0^{-1} 
\exp\left\{ - \frac{1}{4\pi b^2}\int \mbox{d}^2x[\nabla \Phi({\bf
x})]^2\right\}.
\end{equation}

Without loss of generality we can choose  
$\sigma = 1$ in Eq.(\ref{wave})  and 
study the moments of the distribution function of the first 
component of the wave function defined by
\begin{equation}
G(1, ... , N)=
\int D\Phi P[\Phi]\psi^2_1({\bf x}_1)...\psi^2_1({\bf x}_{N}).
\label{defG}
\end{equation}
Notice that all these quantities  are invariant under the shift 
$\Phi\rightarrow\Phi+c$ by an arbitrary real constant $c$. 
The $N$-point moment  of {\it normalizable} wave function 
squares  can always be rewritten as follows:
\begin{equation}
G(1, ... , N)=
\int_0^{\infty}\frac{{\rm d}\mu \exp[- \alpha\mu]\mu^{N- 1}}{(N-1)!}
\int D\Phi\prod_{i = 1}^Ne^{-2\Phi({\bf x}_i)}
e^{-S_{\mu}},
\label{G}
\end{equation}
where the action $S_{\mu}$ is given by
\begin{equation}
S_{\mu}=\int\mbox{d}^2x[\frac{1}{4\pi b^2}(\nabla \Phi)^2+\mu e^{-2\Phi}],
\label{SopenBC}
\end{equation}
and the exponent $\exp[- \alpha\mu]$ is introduced 
to make the integral over $\mu$ 
convergent. Thus, the multi-point moment  (\ref{defG})
is now expressed in terms of the {\it reducible} 
multi-point correlation function 
of LFT. 

Now we shall recall several facts about LFT that we are
going to use. They  can be found, for example,  in Refs.
\cite{seiberg,zamzam}. 
To conform to the notations accepted among the field 
theorists, we rescale the field $\Phi = - b\phi$. 
We also impose the following boundary conditions on the Liouville field:
\begin{equation}
\phi({\bf x})=-Q\ln|{\bf x}|^2,\quad |{\bf x}|\rightarrow\infty,
\quad Q = b +\frac{1}{b}.
\label{BC}
\end{equation}
Thus, we are only considering ground state wave functions 
that decay algebraically at infinity. These are the {\it prelocalized states}. 
If boundary conditions are not chosen properly, the correlation functions
of LFT vanish. In the semiclassical saddle point calculations 
(\cite{falko,seiberg,zamzam}), 
this feature emerges as a condition for the existence of
the saddle point. The boundary condition (\ref{BC}) can be 
formulated as a condition that   the  total flux through the system 
is equal to $Qb$. Uniqueness of the wave functions (1) 
is not violated provided $Qb < 2$ \cite{fradkin}. 

There are different ways to implement the boundary condition (\ref{BC}). 
One can consider the LFT on a large disk $\Gamma$ of radius 
$R \rightarrow \infty$ and add to the local LFT  action 
(\ref{SopenBC}) a boundary term: 
\begin{equation}
S_{\mu,Q} = \frac{1}{4\pi}\int_{\Gamma} \mbox{d}^2x[(\nabla\phi)^2+
4\pi\mu e^{2b\phi}]+
\frac{Q}{\pi R}\int_{\partial\Gamma}\phi \mbox{d}l.
\label{liouv}
\end{equation}
It is more convenient however to  contract the boundary into the point,
thereby transforming the disk into a sphere. The  boundary 
condition is then equivalent to the insertion of the Liouville exponential
$e^{ - 2Q\phi(R)}$ $(R \rightarrow \infty)$ into all
correlation functions \cite{remark}. For $\mu = 0$ one can express
correlation functions of the fields $e^{2\alpha\phi}$ in the LFT with
boundary conditions in terms of correlation functions 
of the conventional Gaussian model:
\begin{eqnarray}
&&
\langle \prod_i e^{2\alpha_i\phi(x_i)}\rangle_Q = \prod_{i < j}
|x_i -
x_j|^{- 4\alpha_i\alpha_j}\delta_{\epsilon}(\sum_i\alpha_i - Q)R^{- 2Q^2},
\nonumber\\
&&
\delta_{\epsilon}(x) = \frac{1}{\pi}\frac{\epsilon}{\epsilon^2 + x^2},
\end{eqnarray}
where $\epsilon$ is a small parameter introduced for regularization
and the subscript $Q$ means that the function on the 
left-hand side is the correlation function calculated with action
(\ref{liouv}). Provided the {\it neutrality condition} $\sum_i\alpha_i = Q$
is satisfied, all correlation functions are proportional to the same
factor $R^{- 2Q^2}$. We shall see that this factor is compensated by
the integral over $\mu$. 
Calculating two-point correlation functions in LFT, we conclude that the
conformal dimension of the Liouville exponential is 
\begin{equation}
\Delta(e^{2\alpha\phi}) = \alpha(Q - \alpha),
\label{dim}
\end{equation}
and this is true for $\mu \neq 0$ also \cite{kpz,d,dk}. 
Note that $Q$ has been chosen such that the exponential operator
in the Liouville action is marginal, 
{\it i.e.,} $\Delta(e^{2b\phi}) = 1$ thus preserving the criticality
of the theory. 

It follows immediately from Eq.(\ref{dim}) that the conformal dimensions 
$\Delta(q)$ ($q$ real) of the composite operators 
$:\psi^{2q}_1({\bf x}): \sim \exp[2qb\phi({\bf x})]$ (the sign $:...:$ denotes
normal ordering) 
are equal to 
\begin{equation}
\Delta(q) = q\left(1 + b^2 - b^2q\right). 
\label{Del} 
\end{equation}
These dimensions depend on the continuous
parameter $b$ which represents the disorder strength.  The LFT 
model remains critical for any values of $b$ 
with the central charge given by 
\begin{equation}
C = 1 + 6Q^2 = 1 + 6(b + 1/b)^2.
\end{equation}
However, at $b^{-2} = M$, where $M = 1, 2, ...$, 
the theory possesses an additional hidden 
SL(2, {\bf R}) symmetry (see, for example, \cite{seiberg}). It is 
also known \cite{zamzam} that three point correlation 
functions have  resonances at these values of $b$.

Eq.(\ref{Del}) reproduces  the dimensions obtained for FRGP in 
Ref.\cite{ludwig} by the replica trick 
and  also  the dimensions obtained for the
conventional localization theory \cite{alt,falko} (in the notation
of Ref. \cite{falko} $b^{-2} = 8\pi^2\beta\nu D$). 
Needless to say that in the conventional 
localization
theory the parameter $b$ undergoes renormalization towards strong
coupling and therefore the
described equivalence holds only in the crossover regime. As we have mentioned earlier, an  extended crossover region exists only 
for the unitary ensemble ($\beta = 2$).

An important application of (\ref{Del}) 
is to the calculation of the scaling with respect to $R$ of the 
inverse participation ratios: 
\begin{equation}
\langle\int_{\Gamma}{\rm d}^2x \psi^{2q}_1({\bf x})\rangle\propto
R^{-\tau(q)},
\quad
\tau(q)=2(1- b^2q)(q-1).
\label{tau}
\end{equation}
The applicability of this scaling for large $q$
has been discussed in Refs. \cite{falko} and \cite{mudry,chamon}. It
was established that $\tau(q)$ must be a monotonously increasing  function of
$q$. This means that Eq.(\ref{tau}) is at most valid for 
$q \leq (1+b^{-2})/2$. 
This condition selects operators 
$\exp(2qb\phi)$ with $qb \leq (b+1/b)/2 = Q/2$. 
Let us note that in
the weak disorder limit $b <<1$, these exponents can still be used for
the description of very high participation ratios.
 
This  selection 
resembles the following well known fact from  the LFT.
Namely, 
there is a difference between operators $\exp(2\alpha\phi)$ with
$\alpha < Q/2$ and $\alpha > Q/2$\cite{seiberg}: only the first ones
(so-called {\it microscopic} operators) 
correspond to local states whereas the latter (so-called {\it macroscopic}
operators) create finite holes on the random surface. 
In LFT  the field $\exp(2b\phi)$ is interpreted as a
metric of a two-dimensional surface. 
In the semiclassical limit $b <<1$  correlation
functions of microscopic operators 
can be obtained by the saddle point approximation. The solution of
classical equations of motion describes a surface spanned on an disk
with radius $R \rightarrow \infty$. This surface has 
constant negative curvature metric with spikes (integrable power law singularities of
the metric field $g_{z\bar z} \sim |z|^{- \eta}, ~ \eta < 2$) at the points
of insertion of operators with $\alpha < Q/2 $ (for $\alpha = Q/2$ we
have punctures, {\it i.e.},
$|z|^{-2}$-singularities). For
macroscopic operators there is no classical solution corresponding to
a surface with a single  boundary at infinity which means that 
each insertion creates a hole of a finite size. 
Since 
the exponents  with $q > (1+ b^{-2})/2$  correspond to macroscopic
operators, it is not correct, in our opinion, to use them for
description of higher order participation ratios. It is important to
mention, however, that macroscopic operators appear in fusion of
microscopic ones.

The authors of Ref. \cite{falko} have stressed 
the importance of the non-universal
prefactor on the right-hand side of (\ref{tau})
to assess the self-consistency of the scaling analysis.
This prefactor is controlled by the dependency on $\mu$ of the
correlation functions of the LFT. Let us note that before 
 we discussed  the results which were independent of
$\mu$, so one can put $\mu =0$.  Now we have to  study
 the effects which arise in the full theory
 with nonzero $\mu$.

It is easy to find the scale ($\mu$) dependence of any
correlation function in Liouville theory \cite{kpz,d,dk}:
\begin{eqnarray}
&&
\tilde G(1, ... , N) = 
\langle \prod_{i=1}^{N} e^{2\alpha_{i}\phi({\bf x}_i)}\rangle_Q =
\nonumber\\
&&
(\pi\mu)^{(Q - \sum_{i=1}^N \alpha_i)/b} 
F_{\alpha_{1}...\alpha_{N}}({\bf x}_{1}, \cdots,{\bf x}_{N}).
\label{tildeG}
\end{eqnarray}
Here
$F_{\alpha_{1}...\alpha_{N}}({\bf x}_{1},...,{\bf x}_{N})$
must be calculated in the Liouville theory with $\mu =1$ and the
answer for the generic correlation function is unknown. For three- and
four-point functions the answer was found in \cite{do,zamzam}
for any $b$.
However for the special case when the factor 
$(Q - \sum_i \alpha_i)/b$ is an integer, {\it i.e.,} when the
correlation function (\ref{tildeG})
is proportional to an integer power of $\mu$,
all multi-point correlation functions  can be obtained explicitly by expanding 
$\exp(-S_{\mu,Q})$ [cf. (\ref{liouv})] in powers of $\mu$. 

Thus, unlike the case of conformal dimensions where Eq.(\ref{Del})
holds for any $b$, the multi-point correlation functions
can be easily obtained  only for discrete values of the disorder strength
$b^2  = 1/M, \: M =1, 2, ... $,
for which the expansion over $\mu$ contains only {\it one}
non-vanishing term. 
Indeed when $b^{-2}$ is an integer, 
the multi-point correlation functions of the 
Liouville theory can be expressed in terms of correlation functions  
of the free bosonic field
(the Dotsenko-Fateev construction \cite{fateev}).
To see this, we recall that for the correlation function (\ref{G})
all $\alpha_i = b$. Hence,
\begin{eqnarray}
(Q - \sum^{N}_{i=1} \alpha_i)/b =  1 - N + 1/b^2. \label{neut}
\end{eqnarray}
Provided $b^{-2} = M$ with integer $M\geq N-1$ the neutrality condition 
is fulfilled with the term $\sim \mu^{1 - N + M}$. 
The result for the $N$-point function on the 
infinite plane follows immediately:
\begin{eqnarray}
&&
\tilde G(1, ... ,N) = 
\epsilon^{-1}R^{- 2Q^2}\frac{\mu^{1 - N + M}}{(1 - N + M)!}
\prod_{i < j}^N\frac{1}{|z_{ij}|^{4/M}}
\int \left(\prod_{l =1}^{1 - N + M}\mbox{d}^2\xi_l\right) 
\nonumber\\
&&
\times
\left(
\prod_{l = 1}^{1 - N + M}
\prod_{i = 1}^N
\frac{1}{|z_i - \xi_l|^{4/M}}\right)
\left(\prod_{m < n}^{1 - N + M}
\frac{1}{|\xi_{mn}|^{4/M}} 
\right)
\label{g}
\end{eqnarray}
(here $z, \bar z = x
\pm  $i$y$ are complex coordinates).  The expression for a finite
system can be obtained by conformal transformation according to the
general rules of conformal theory. Substituting Eq.(\ref{g})  into
Eq.(\ref{G}) we find that all correlation functions are proportional
to the same nonuniversal  factor 
$D = \epsilon^{-1}R^{- 2Q^2}\int^{\infty}_{0} 
\mbox{d}\mu \mu^{1/b^2}e^{-\alpha\mu}$.
We remark that by regularizing the theory with the introduction 
of $\exp[-\alpha\mu]$, we overestimate the normalization
factors of the wave functions. In other words,
we restrict the allowed spatial fluctuations of the wave function
amplitude from above, and thereby eliminate the contributions from
localized states to the statistical average in (\ref{defG}).
Finally we note that  as shown 
in Ref. \cite{chamon}, $b = 1$  
is the largest value of the disorder strength for which 
quenched and annealed averages 
[with respect to the Gaussian distribution (2)] 
of the normalization 
$\int{\rm d}^2x \exp[-2\Phi({\bf x})]$
aggree. On the other hand, $b = 1$ corresponds to the
minimum of the Liouville central charge $C_{min} = 25$. We recall that 
when  $b = 1$ the total flux through the system reaches the value of 2 
which corresponds to the (first) change in the ground state degeneracy. 

We conclude with a remark about a difference between even and odd $M$
in Eq.(\ref{g}).  For even $M$, the 
operator with maximal dimension $\Delta_{max} = 1/2 + M/4$ 
is $O_{M/2}(x) \equiv \exp( Mb\phi)$. 
>From Eq.(\ref{g}) we find:
\begin{eqnarray}
\langle O_{M/2}(z_1)O_{M/2}(z_2)\rangle &=& 
\frac{D}{|z_{12}|^{M}}\int \frac{d^2\xi}{|z_1 - \xi|^2|z_2 - \xi|^2} 
\nonumber\\
&\approx&
\frac{4D}{|z_{12}|^{M + 2}}\ln\left(|z_{12}|/a\right).
\label{dd}
\end{eqnarray}
Thus, the two-point correlation function of this operator contains 
a logarithm. This does not happen when $M$ is odd. 
Indeed, the operator with maximal dimension $\Delta_{max} = (M + 1)^2/4M$ is
$O_{(M + 1)/2}(x)$ and its two-point correlation function has 
the conventional form.  
It may be that, as in other theories with logarithms 
(see Gurarie \cite{gur} and Caux {et al.}\cite{caux}), 
Eq. (\ref{dd}) indicates the presence of an additional operator $C(x)$ 
in the operator algebra of the theory such that
\begin{eqnarray}
&&
\langle C(z_1)O_{M/2}(z_2)\rangle = - \frac{4D}{|z_{12}|^{M + 2}},
\nonumber\\
&&
\langle C(z_1)C(z_2)\rangle = 0 .
\label{C}
\end{eqnarray}
We are not certain however 
that the same interpretation is valid for the Liouville theory 
where there is no one-to-one correspondence between operators and 
states and one therefore must be careful formulating the operator expansion. 
It is also interesting to note that for $b= M =1$  
we have the Liouville theory  with central charge $C_{min} = 25$
which is known to have 
logarithmic operators with scaling dimensions one \cite{bilal}. 
As we discussed earlier,
when  $b = 1$ the total flux through the system reaches the value of 2 
which corresponds to the appearence of a second ground state in our
model. The relationship between the existence of
marginal logarithmic operator and the change of the ground state degeneracy
is an interesting open question.

 In more complicated models where the disorder depends on several
fields we expect that the corresponding universality classes will be
connected with $W_n$-gravities. Another interesting problem is to find
what deformation of the Liouville theory leads to
localization and thus to reproduce the renormalization group equations
obtained by Altshuler {\it et al.}\cite{alt}. 

 We are grateful to J. Chalker, A. Georges, M. Gunn, 
K. Efetov, M. P. A. Fisher, 
D. Khveshchenko,   
V. Kravtsov, D. Khmelnitskii, I. Lerner, A. W. W. Ludwig, 
A. A. Nersesyan and especially to V. Fal'ko and E. Fradkin 
for illuminating discussions, constructive criticism  
and interest for the work. 
C.M. acknowledges financial support from the Swiss National funds.


\begin{thebibliography}{99}
\bibitem{alter} Alternatively,  
this problem can be formulated 
as the  problem of  a quantum particle  with  the quadratic dispersion law 
$\epsilon({\bf k}) = k_x^2 + k_y^2$ 
in the following potential:
$V({\bf x}) = [\nabla \Phi({\bf x})]^2 - \nabla^2\Phi({\bf x}).$
\bibitem{remark} It is possible to rewrite this boundary term
as an integral  $(Q/4\pi)\int \mbox{d}^2 x \sqrt{{\cal G}}{\cal R}$, where
$\cal{G}$ and $\cal{R}$ are some background metric and curvature.
Choosing  background which is flat except for one  singular point at 
infinity, one can obtain the previous result.
\end{thebibliography}
\end{document}